\documentstyle[epsf,12pt]{article}
\newcommand{\newsection}{    
\setcounter{equation}{0}
\section}
\def\appendix#1{
\addtocounter{section}{1}
\setcounter{equation}{0}
\renewcommand{\thesection}{\Alph{section}}
\section*{Appendix \thesection\protect\indent #1}
\addcontentsline{toc}{section}{Appendix \thesection\ \ \ #1}
}
\newcommand{\rf}[1]{(\ref{#1})}

\def\be{\begin{equation}}
\def\ee{\end{equation}}
\newcommand{\beq}{\begin{equation}}
\newcommand{\eeq}{\end{equation}}
\newcommand{\bea}{\begin{eqnarray}}
\newcommand{\eea}{\end{eqnarray}}

\renewcommand{\l}{\lambda}

\newcommand{\om}{\omega}

\newcommand{\non}{\nonumber}

\newcommand{\tr}{{\,\rm tr}\:}

\begin{document}
\topmargin 0pt
\oddsidemargin 5mm
\headheight 0pt
\headsep 0pt
\topskip 9mm

\hfill NORDITA-96/28P

\hfill SMI-Th--96/13

\addtolength{\baselineskip}{0.20\baselineskip}
\begin{center}
\vspace{26pt}
{\large \bf {Hermitian Matrix Model with Plaquette Interaction }}
\newline
\vspace{26pt}

{\sl L.\ Chekhov} \footnote{{Supported by RFFI Grant 
No.\,96--01--00344}}\\ 
\vspace{6pt} 
Steklov Mathematical Institute\\ Vavilov 
st.\ 42, 117966, GSP-1, Moscow, Russia\\

\vspace{18pt}
{\sl C. Kristjansen}\\ 
\vspace{6pt}
NORDITA \\
 Blegdamsvej 17,
DK-2100 Copenhagen \O, Denmark \\
\end{center}
\vspace{20pt} 
\begin{center}
{\bf Abstract}
\end{center}

\noindent
We study a hermitian $(n+1)$-matrix model with plaquette interaction,
$\sum_{i=1}^n MA_iMA_i$. By means of a conformal transformation we
rewrite the model as an $O(n)$ model on a random lattice with a
non polynomial potential. This allows us to solve the model
exactly. We
investigate the critical properties of the plaquette model and find
that for $n\in]-2,2]$ the model belongs to the same universality class
as the $O(n)$ model on a random lattice.
\newpage

\newsection{Introduction}
Despite the fact that great progress has been made in solving matrix models
in recent years many interesting models remain unsolved. One important
class of models for which an exact solution is still lacking is models
with ``plaquette type'' interactions. Lattice gauge theories like the
Weingarten model~\cite{Wei80,EK82} and the Kazakov-Migdal model~\cite{KM93} are
typical examples of such models but recently also plaquette type
models without gauge degrees of freedom have attracted attention,
namely as generating functionals for Meander numbers~\cite{DGG95}.
In the present paper we will consider the following model 
\beq
Z=e^{N^2F}=\int d M \prod_{i=1}^n dA_i \exp\left\{-N\tr\left(V(M)
+\frac{1}{2}\sum_{i=1}^nA_i^2-\frac{1}{2}\sum_{i=1}^nMA_iMA_i\right)\right\}
\label{definition}
\eeq
where all the matrices are hermitian and $V(M)$ is an arbitrary
polynomial potential. When $n=1$ this model shows a large degree of
similarity with the 2-dimensional reduced Weingarten
model~\cite{Wei80,EK82} which is given by
\beq
Z=\int \prod_{\mu=1}^2 dA_{\mu}^{\dagger} dA_{\mu}\exp
\left\{-N\tr\left[\sum_{\mu=1}^2A_{\mu}^{\dagger}A_{\mu}-
g\sum_{\mu,\nu=1}^2\left(A_{\mu}A_{\nu}A_{\mu}^{\dagger}A_{\nu}^{\dagger}
+h.c.\right)\right]\right\}.
\eeq
However, the two models are not equivalent. A model equivalent
to~\rf{definition} for $n=1$ involving complex matrices is 
\[
Z=\int dM_1 dM_2dA^{\dagger}dA 
\exp\left\{-N\tr\left[V(M_1)+\frac{1}{2}M_2^2
+A^{\dagger}A- M_1A^{\dagger}M_2A\right]
\right\} \nonumber
\]
where $M_1$ and $M_2$ are hermitian and $A$ is complex. Our
model~\rf{definition} also shows some similarity with matrix models generating
Meander numbers~\cite{DGG95}.
 Its interaction is of the type needed for such
models. However, our model is too simple to provide a
generating functional for Meander numbers. For that purpose one must
be able to work also with an arbitrary number of $M$-matrices. Let us
finish by mentioning that  our solution of the model~\rf{definition} gives
the solution to a simple three-matrix problem, namely the following
\beq
Z=\int dA\, dB\, dC\exp\left\{-N\tr \left(V(A)+\frac{1}{2}B^2+
\frac{1}{2}C^2-g ABC\right)\right\}.
\label{three}
\eeq
The partition function~\rf{three} can be brought on the
form~\rf{definition} (with $n=1$) by integrating out one of the 
three matrices. In reference~\cite{Cic93} the model~\rf{three} with 
$V(A)=\frac{1}{2}A^2$ was studied numerically.

The paper is organized as follows. In section~\ref{saddle} we derive
the saddle point equation of the model~\rf{definition} and argue that
it has the same structure as that of the $O(n)$ model on a random
lattice~\cite{Kos89}. Then in section~\ref{O(n)} we explicitly 
transform the
model into an $O(n)$ model with a somewhat unconventional
potential. Exploiting the already known exact solution of the $O(n)$
model on a random lattice~\cite{EK95}, we hereafter in section~\ref{solution}
write down the solution of  the present model. In section~\ref{gaussian}
we specialize to a quadratic potential and perform a detailed analysis of this 
case. In particular we investigate the critical properties of the model and
find that for $n\in ]-2,2]$ the model~\rf{definition}
belongs to the same universality class as the ordinary $O(n)$ model on
a random lattice.  Section~\ref{conclusion} contains our conclusion
and outlook. Finally in an appendix we comment on the Virasoro algebra
structure carried by our model.

\newsection{The saddle point equation \label{saddle} }

Let us carry out the gaussian integration over the $A$-matrices 
in~\rf{definition}. This gives 
\beq
Z= \int dM\exp\left\{-N\tr V(M)\right\} \det\left(I\otimes I-
M^{T}\otimes M\right)^{-n/2}.
\label{Aint}
\eeq
where $M^{T}$ is the transpose of $M$. Next, let us diagonalize the
$M$-matrices and integrate out the angular degrees of freedom. This
leaves us with the
following integral over the eigenvalues, $\{\lambda_i\}$, of the matrix
$M$
\beq
Z\propto \int \prod_i d\l_i\exp\left\{-N\sum_jV\left(\lambda_j\right)\right\}
\prod_{j<k}(\l_j-\l_k)^2\prod_{j,k}(1-\l_j\l_k)^{-n/2}.
\label{eigenvalue}
\eeq
In the limit $N\rightarrow \infty$ the eigenvalue configuration is
determined by the saddle point of the integral above~\cite{BIPZ78}.
The corresponding saddle point equation reads
\beq
NV'\left(\lambda_i\right)=2\sum_{j\neq i}\frac{1}{\lambda_i-\lambda_j}+
n\sum_j\frac{\l_j}{1-\l_j\l_i}
\label{saddle1}
\eeq
or
\beq
V'\left(\lambda_i\right)+\frac{n}{\l_i}=
\frac{2}{N}\sum_{j\neq i}\frac{1}{\lambda_i-\lambda_j}
+\frac{n}{N}\frac{1}{\l_i^2}\sum_j\frac{1}{\frac{1}{\lambda_i}-\lambda_j}
\label{saddle2}.
\eeq
Following~\cite{BIPZ78} we now introduce an eigenvalue
density
$\rho(\lambda)=\frac{1}{N}\sum_i\delta\left(\lambda-\lambda_i\right)$
which in the limit $N\rightarrow \infty$ becomes a continuous
function.
As is clear from equation~\rf{eigenvalue} the model becomes singular
if one of the eigenvalues approaches $+1$ or $-1$. We shall hence
solve the model with the requirement that the support of the
eigenvalue distribution does not include these points. To be precise
we will assume that the eigenvalues are confined to one interval,
$[\alpha,\beta]$ , $-1<\alpha\leq \beta<1$ and that the
corresponding eigenvalue distribution is normalized to one. Using the
hence obtained
solution we will afterwards investigate what happens when, say,
$\alpha$ approaches 1. Again following~\cite{BIPZ78} we
introduce the one-loop correlator $\Omega(p)$ by
\beq
\Omega(p)=\int_a^b d\mu\, \frac{\rho(\mu)}{p-\mu}.
\eeq
In terms of the one-loop correlator the saddle point
equation~\rf{saddle2} can be written as
\beq
V'(p)+\frac{n}{p}=\Omega(p+i0)+\Omega(p-i0)+n\,\frac{1}{p^2}\,
\Omega(\frac{1}{p}),
\hspace{0.7cm} p\in [\alpha,\beta]
\eeq
or with $\overline{V}\,'(p)= pV'(p)+n$ and $\overline{\Omega}(p)=p\,\Omega(p)$
\beq
\overline{\Omega}(p+i0)+\overline{\Omega}(p-i0)
+n\overline{\Omega}(\frac{1}{p})=\overline{V}\,'(p),
\hspace{0.7cm}p\in [\alpha,\beta].
\eeq
This equation, in analogy with the saddle point equation
of the $O(n)$ model on a random lattice, involves two cuts. First
there is the cut of the function $\overline{\Omega}(p)$. This
cut is the physical cut, i.e.\ the support of the eigenvalue
distribution corresponding to the matrix $M$. In addition to the
physical cut another cut turns up in the saddle point equation, namely
the cut of the function $\overline{\Omega}(\frac{1}{p})$. The singular
behaviour referred to above corresponds to the situation where the two
cuts merge.
\newsection{Transformation to $O(n)$ model \label{O(n)} }
In order to fully exploit the similarity of our model with the $O(n)$
model on a random lattice we will explicitly bring it on 
the $O(n)$ model form.
 For that purpose, let us perform the
following redefinitions of our matrix fields
\beq
M \rightarrow \frac{1-X}{1+X}, \hspace{1.0cm}
A_i \rightarrow \frac{1}{2}\left(1+X\right)^{1/2}S_i\left(1+X\right)^{1/2}
\label{AXdef}
\eeq
i.e.,
\beq
dM \rightarrow \left[\det(1+X)\right]^{-2N}dX,
\hspace{1.0cm} dA_i \rightarrow \left[\det(1+X)\right]^NdS_i.
\eeq
Inserting these expressions in our original partition function we get
\beq
Z\propto \int dX\prod_{i=1}^ndS_i\exp\left\{-N\tr\left(
V\left(\frac{1-X}{1+X}\right)+(2-n)\log(1+X)+X\sum_{i=1}^nS_i^2\right)\right\}.
\eeq
This model is nothing but the $O(n)$ model on a random lattice with
the somewhat unconventional potential 
\beq
U(p)=V\left(\frac{1-p}{1+p}\right)+(2-n)\log(1+p)\equiv
U_0(p)+(2-n)\log(1+p).
\label{Onpot}
\eeq
The saddle point equation for this model reads~\cite{Kos89}
\beq
W(p+i0)+W(p-i0)+nW(-p)=U'(p),
\label{saddle3}
\eeq
where $W(p)$ is the one-loop correlator of the $X$-field. The physical
cut now extends from $a=(1+\alpha)/(1-\alpha)$ to
$b=(1+\beta)/(1-\beta)$
and the unphysical cut is the mirror image with respect to zero of the
physical cut. We note that the point $p=1$ lies on the physical cut
and the point $p=-1$ on the unphysical cut. As before we expect some
kind of singularity to occur when the physical and the unphysical cut
merge and we will have to assume that $a>0$. Since the potential
includes a logarithmic term we might also expect a Penner like
singularity to appear, i.e.\ a singularity corresponding to the
physical and the unphysical cut degenerating to respectively the
 point $p=+1$ and the point 
$p=-1$~\cite{Penner}.
\newsection{The solution \label{solution}}
In~\cite{EK95} an exact contour integral representation of the
1-loop correlator of the $O(n)$ model on a random lattice was written
down. The derivation of the exact formula was based on the assumption
of the potential of the model being polynomial. However, it is easy to
convince oneself that the formulas remain valid (when written in the
appropriate way) as long as the potential or rather its derivative
does not have any singularities which intervene with the physical cut
of the one-loop correlator. For our model the only singular point of
$U'(p)$ is $p=-1$ (cf.\ to equation~\rf{Onpot}) and, as argued
earlier, this point always lies on the unphysical cut. Hence we can
take over the solution of the $O(n)$ model on a random lattice 
from~\cite{EK95}. Let us remind the reader of the structure of the
solution. First, it is convenient to decompose the 1-loop correlator into
a regular part, $W_r(p)$, having no cut, and a singular part, $W_s(p)$
\beq
W(p)=W_r(p)-W_s(p).
\eeq
It follows from equation~\rf{saddle3} that $W_r(p)$ is given by
\beq
W_r(p)=\frac{2U'(p)-nU'(-p)}{4-n^2}
\eeq
while $W_s(p)$ is a solution to the homogeneous saddle point
equation. As shown in reference~\cite{EK95} any solution of the
homogeneous saddle point equation can be parametrized in terms of two
auxiliary functions, $G(p)$ and $\tilde{G}(p)$. More precisely, any such
solution, $S(p)$, can be written as
\beq
S(p)={\cal A}(p)G(p)+{\cal B}(p)p\tilde{G}(p)
\label{parametrization}
\eeq
where ${\cal A}(p)$ and ${\cal B}(p)$ are regular but not necessarily
entire functions. The function $G(p)$ is defined by the following
three requirements
\begin{enumerate}
\item
$G(p)$ is a solution of the homogeneous saddle point equation
corresponding to $n=2\cos(\nu \pi)$, i.e.,
$$
G(p+i0)+G(p-i0)+nG(-p)=0.
$$
\item
$G(p)$ is analytic in the complex plane except for the cut $[a,b]$ and
behaves as $(p-a)^{-1/2}$ and $(p-b)^{-1/2}$ in the vicinity of $a$
and $b$.
\item
$G(p)$ has the following asymptotic behaviour
\[G(p)\sim \frac{1}{2\cos(\nu\pi/2)}\,\frac{1}{p}, \hspace{0.7cm}
p\rightarrow \infty.\]
\end{enumerate}
These three requirements are enough to determine $G(p)$ uniquely and a
completely explicit expression for $G(p)$ in terms of theta-functions
can be written down~\cite{EK95A}. We shall not need the detailed form of
$G(p)$ for the following but let us mention that, as is obvious from
the definition, $G(p)$ does not contain any explicit reference to the
matrix model coupling constants. Furthermore the dependence of $G(p)$
on $n$ appears explicitly only via  a parameter $e$ given by
\beq
e=a\,\mbox{sn}\,\left(i(1-\nu) K',k\right), \hspace{0.7cm}k=\frac{a}{b}.
\eeq
The function $\tilde{G}(p)$ is defined in a way analogously to
$G(p)$. Only $\nu$ is replaced by $1-\nu$. Hence $\tilde{G}(p)$ is a
solution of the homogeneous saddle point equation with $n$ replaced by
$-n$. Now, if $\tilde{G}(p)$ is a solution of the homogeneous saddle
point equation corresponding to $-n$ then obviously $p\tilde{G}(p)$ is
a solution to the original saddle point equation. This explains the
appearance of the combination $p\tilde{G}(p)$ in
relation~\rf{parametrization}.
In a compact form the full one-loop correlator, $W(p)$, can be written
as (cf.\ to~\cite{EK95})
\bea
W(p)&=& \frac{1}{4-n^2}\left\{ ip\,\tilde{G}(p)\oint_C \frac{d\omega}{2\pi
i} \frac{U'(\omega)}{p^2-\om^2}\left[\left(p^2-e^2\right) i
\tilde{G}(\om)+\frac{\overline{\sqrt{e}}}{e}\om G(\om)\right] \right. 
\nonumber \\
&& \left.-G(p)\oint_C\frac{d\om}{2\pi i}\frac{U'(\om)}{p^2-\om^2}\left[
\left(p^2-\tilde{e}^2\right)\om G(\om)
+p^2\frac{\overline{\sqrt{\tilde{e}}}}{\tilde{e}}i\tilde{G}(\om)\right]\right\}
\label{contour}
\eea
where the contour $C$ encircles the physical cut $[a,b]$ but not the
points $\om=\pm p$ and where $\overline{\sqrt{e}}$ is defined by
\beq
\overline{\sqrt{e}}=\sqrt{\left(e^2-a^2\right)\left(e^2-b^2\right)}
=-ab\,\mbox{cn}\left(i(1-\nu)K'\right)\mbox{dn}\left(i(1-\nu)K'\right).
\eeq
Here and in the following we will use the convention that tilded
quantities appear from untilded ones by the replacement
$\nu\rightarrow 1-\nu$. If one wants to evaluate the contour
integral~\rf{contour} in a specific case, the most convenient line of
action is to deform the contour into several different contours encircling
respectively the points $\om=\pm p$ and the various singularities of
$U'(p)$. The contribution from the poles $\om=\pm p$ then
gives rise to the regular part of $W(p)$ while the contribution from
singularities of $U'(p)$ gives the singular part of $W(p)$. The
expression~\rf{contour} must be supplemented by a set of boundary equations
which determine the endpoints of the physical cut, $a$ and $b$. These
equations read
\beq
\oint_{C}\frac{d\om}{2\pi i}\,V'(\om)\tilde{G}(\om)=0, 
\label{B1}
\eeq
\beq
\oint_{C}\frac{d\om}{2\pi i}\om V'(\om){G}(\om)=
\frac{2-n}{\sqrt{1+\frac{n}{2}}}
\label{B2}
\eeq 
and ensure the correct asymptotic behaviour of the one-loop
correlator,
 namely $W(p)\sim 1/p$ as $p\rightarrow \infty$. In~\cite{EK95} 
 it was shown that for the ordinary $O(n)$ model
on a random lattice the higher genera contributions to the correlators
and the free energy simplify considerably if one expresses the $p$-dependence
via a set of basis functions
$\left\{G_a^{(k)}(p),G_b^{(k)}(p)\right\}$ and the dependence on the
coupling constants via a set of moment variables
$\left\{M_k,J_k\right\}$. Needless to say that a similar
simplification can be obtained in the present case.

\newsection{The quadratic potential \label{gaussian} } 
For simplicity, let us now restrict ourselves to the case where the
potential $V(M)$ in~\rf{definition} is given by
\beq
V(M)=\frac{1}{2T}M^2.
\eeq
The analysis of the general case can be done along the same lines.  
For $U(p)$ in equation~\rf{Onpot} we then obviously have
\beq
U(p)=\frac{1}{2T}\left(\frac{1-p}{1+p}\right)^2+(2-n)\log(1+p).
\label{Ufin}
\eeq
\subsection{The boundary equations}
Inserting~\rf{Ufin} into the boundary equation~\rf{B1} we get
\beq
(2-n)\tilde{G}(-1)-\left.\frac{2}{c}\left[\frac{\partial^2}{\partial
p^2}-\frac{\partial}{\partial p}\right]
\tilde{G}(p)\right|_{p=-1}=0. 
\label{B11}
\eeq
Here the first term comes from the pole at $w=-1$ in the
logarithmic term and the second from the pole at $\om =-1$ in
$U_0'(p)$. There is no contribution from infinity. Next inserting the
expression for $U'(p)$ into~\rf{B2} we get
\beq
(2-n)G(-1)+\left.\frac{6}{c}\frac{\partial}{\partial 
p}G(p)\right|_{p=-1} -\left.\frac{2}{c}\left[\frac{\partial^2}{\partial 
p^2}+1\right] G(p)\right|_{p=-1}=0, 
\label{B22} 
\eeq 
where the first term comes 
from the pole at $\om=-1$ in the logarithmic term and the two next 
from the pole at $\om=-1$ in $U_0'(p)$. In this case we do have a 
contribution from infinity but it cancels with the constant on the right 
hand side of the original equation. 
To proceed we need to know
 $\frac{\partial}{\partial p} 
G(p)$ and $\frac{\partial^2}{\partial p^2}G(p)$. These can be found by
exploiting the fact that any solution to the homogeneous saddle point
equation corresponding to $n=2\cos(\nu\pi)$ has a parametrization of
the form~\rf{parametrization} (and similarly for the saddle point
equation corresponding to $n=2\cos\left((1-\nu)\pi\right)$). The exact
form of the parametrization is determined by the requirements on the
analyticity properties and the asymptotic behaviour of the functions
in question (cf.\ to~\cite{EK95}). The result reads
\beq
\frac{\partial}{\partial p}G(p)=\frac{1}{(p^2-a^2)(p^2-b^2)}\left\{
p\left(e^2-p^2-\alpha\frac{\overline{\sqrt{e}}}{e}\right)G(p)
+i\left(\alpha
p^2+e^2\tilde{\alpha}\right)\tilde{G}(p)\right\},
\label{deriv1}
\eeq
\bea
\lefteqn{
\frac{\partial^2}{\partial p^2}G(p)=\frac{1}{(p^2-a^2)^2(p^2-b^2)^2}\times 
\left\{\left[2p^6+\left(a^2+b^2-
5\left(e^2-\alpha\frac{\overline{\sqrt{e}}}{e}\right)
-\alpha \tilde{\alpha}\right)p^4 \right. \right.} \non \\
&&\non \\
&&\left.+\left(\left(a^2+b^2\right)\left(2e^2-\alpha\frac{\overline{\sqrt{e}}}{e
}
-\alpha^2\right)-4a^2b^2\right)p^2
+a^2b^2\left(e^2-\alpha\tilde{\alpha}-\alpha\frac{\overline{\sqrt{e}}}{e}\right)
\right] G(p) \non \\
&& \non \\
&&+\left.\left[4\alpha p^4+\left(6e^2\tilde{\alpha}-\alpha(a^2+b^2)\right)p^2
-3e^2\tilde{\alpha}\left(a^2+b^2\right)-2\alpha a^2b^2\right](-i)p\tilde{G}(p)
\right\}
\label{deriv2}
\eea
where 
\beq
\alpha=b\left(Z\left(i(1-\nu)
K',k\right)+i(1-\nu)\frac{\pi}{2K}\right),
\hspace{0.7cm}k=\frac{a}{b}.
\eeq
Exploiting the explicit expression for $G(p)$ found in~\cite{EK95A} 
one can derive the following useful relation
between $G(-1)$ and $\tilde{G}(-1)$
\bea
\tilde{G}(-1)&=&-i\sqrt{k}\,\mbox{sn}(i\nu K',k)G(-1)
\hspace{0.5cm}  \nonumber \\
&=&-i\,\left(\sqrt{ab}\,\right)^{-1}\tilde{e}\, G(-1). \label{GG}
\eea
Now, let us for a moment go back to our original
model~\rf{definition}. With a quadratic potential the model is
invariant under the transformation $M\rightarrow -M$ and the
eigenvalues of the matrix $M$ must hence live on an interval of the
type $[-\alpha,\alpha]$. This means that for the support $[a,b]$ of the
eigenvalue distribution of the matrix $X$, defined in~\rf{AXdef}, we
have $b=\frac{1}{a}$. 
Exploiting~\rf{GG} and setting $b=1/a$ and we get from~\rf{deriv1}
and~\rf{deriv2} 
\beq
\left.\frac{\partial}{\partial p}G(p)\right|_{p=-1}=
-\frac{a^2}{(1-a^2)^2}
\left\{\left(\alpha\frac{\overline{\sqrt{e}}}{e}-e^2+1\right)
+\left(\alpha\tilde{e}-e\tilde{\alpha}\right)\right\}G(-1),
\label{deriv1f}
\eeq
\beq
\left.\frac{\partial^2}{\partial p^2}G(p)\right|_{p=-1}=
\frac{a^2}{(1-a^2)^2}\left\{
1-\alpha\tilde{e}+3e\tilde{\alpha}+2e^2-\alpha^2-
\alpha\frac{\overline{\sqrt{e}}}{e}\right\}G(-1).
\label{deriv2f}
\eeq
Finally
inserting~\rf{deriv1f} and~\rf{deriv2f} into~\rf{B11} and~\rf{B22} both
 equations
reduce to
\beq
(2-n)-\frac{2a^2}{(1-a^2)^2}\,\frac{1}{T}\left\{
2+2\tilde{e}{\alpha}+\tilde{e}^2-\tilde{\alpha}^2\right\}=0.
\label{Bfin}
\eeq
\subsection{The string susceptibility}
In this section we will determine the quantity
$\frac{d}{dT}T^3\frac{dF}{dT}$ which we will make use of later when
investigating the critical behaviour of the model. Here $F$ stands for
the genus zero contribution to the free energy of our model. The
quantity $\frac{d}{dT}T^3\frac{dF}{dT}$ is related to the string
susceptibility $U(T)=\frac{d^2}{dT^2}\left(T^2 F\right)$ by
\beq
\frac{d^2}{dT^2}\left(T^3\frac{dF}{dT}\right)=T\frac{dU(T)}{dT}.
\label{U(T)}
\eeq
By direct computation we find
\beq
T^2\frac{dF}{dT}=\frac{1}{2}\left\langle
\frac{1}{N}\tr\left(\frac{1-X}{1+X}\right)^2\right\rangle
=\frac{1}{2}\oint_{C_1}\frac{d\om}{2\pi
i}\left(\frac{1-\om}{1+\om}\right)^2 W(\om).
\eeq
Multiplying by $T$ and differentiating once more
gives
\beq
\frac{d}{dT}\,T^3\,\frac{dF}{dT}=\frac{1}{2}\oint_{C_1}\frac{d\om}{2\pi
i}\left(\frac{1-\om}{1+\om}\right)^2\frac{d}{dT}\left(TW(\om)\right)
\label{secondderivative}
\eeq
Now, it follows from~\rf{saddle3} that $\frac{d}{dT}\left(TW(p)\right)$ fulfills 
the
following equation
\beq
\frac{d}{dT}\left(TW(p+i0)\right)+\frac{d}{dT}\left(TW(p-i0)\right)+
n\frac{d}{dT}(TW(-p))=\hat{V}'(p)
\label{saddlederiv}
\eeq
where
\beq
\hat{V}'(p)=(2-n)\frac{1}{1+p}.
\eeq
{}Furthermore we obviously have for the asymptotic behaviour
\beq
\frac{d}{dT}\left(TW(p)\right)\sim
\frac{1}{p},\hspace{0.5cm}\mbox{as}\hspace{0.5cm} p\rightarrow \infty
\label{asymptotic}
\eeq
and as regards the analyticity structure, $\frac{d}{dT}\left(TW(p)\right)$
must be analytic in the complex plane outside the support of the
eigenvalue distribution and behave as
\beq
\frac{d}{dT}\left(TW(p)\right)\sim (p-a)^{-1/2},\,\,(p-b)^{-1/2}
\hspace{0.5cm}\mbox{for}\hspace{0.5cm}p\rightarrow a,b
\eeq
Let us introduce the following notation
\beq
W_T(p)\equiv \frac{d}{dT}\left(TW(p)\right)
\eeq
and let us split $W_T(p)$ in a regular part, $W_T^r(p)$, and a singular
part, $W_T^s(p)$, i.e.\
\beq
W_T(p)=W_T^r(p)-W_T^s(p)
\eeq
where $W_T^r(p)$ does not have any cut.
 Then we have from~\rf{saddlederiv} 
\beq
W_T^r(p)=\frac{2\hat{V}'(p)-n\hat{V}'(-p)}{4-n^2}=\frac{1}{1-p^2}\left\{
\frac{2-n}{2+n}-p\right\}.
\label{regular}
\eeq
The singular part of $W_T(p)$ is a solution of the homogeneous saddle
point equation and as any other such solution has a parametrization of
the form~\rf{parametrization}. 
Since $W_T^r(p)$ has poles at $p=\pm1$,
$W_T^s(p)$ must likewise have poles here because the full function
$W_T(p)$ should be analytic outside the support of the eigenvalue
distribution.
Therefore we can write
\beq
W_T^s(p)=\frac{1}{1-p^2}\left\{A(p^2)G(p)+
pB(p^2)\tilde{G}(p)\right\}
\eeq
where $A(p^2)$ and $B(p)^2$ are now entire functions. From the
requirement~\rf{asymptotic} on the asymptotic behaviour and the
expression~\rf{regular} for $W_T^r(p)$ one can conclude
that $A(p^2)$ and $B(p^2)$ must
be constants. Hence we have
\beq
W_T(p)=\frac{1}{1-p^2}\left\{\frac{2-n}{2+n}-p
+AG(p)+pB\tilde{G}(p)\right\}
\eeq
and the constants $A$ and $B$ are determined by the requirement that
the poles at $p=\pm 1$ should vanish, i.e.\
\beq
\frac{2-n}{2+n}-1+AG(1)+B\tilde{G}(1)=0,\\
\eeq
\beq
\frac{2-n}{2+n}+1+AG(-1)-B\tilde{G}(-1)=0.
\eeq
The solution reads
\beq
A=\frac{1}{2+n}\left\{\frac{2n\tilde{G}(-1)-4\tilde{G}(1)}
{G(1)\tilde{G}(-1)+G(-1)\tilde{G}(1)}\right\},
\label{A}
\eeq
\beq
B=\frac{1}{2+n}\left\{\frac{2n{G}(-1)+4{G}(1)}
{G(1)\tilde{G}(-1)+G(-1)\tilde{G}(1)}\right\}.
\label{B}
\eeq
Going back to~\rf{secondderivative} we can write
\bea
\frac{d}{dT}T^3\frac{dF}{dT}&=&\frac{1}{2}\oint_{C_1}\frac{d\om}{2\pi i}
\left(\frac{1-\om}{1+\om}\right)^2\frac{1}{1-\om^2}\left\{AG(\om)
+\om B \tilde{G}(\om)\right\}\nonumber \\
&& \nonumber \\
&=&\frac{1}{2}\oint_{C_1}\frac{d\om}{2\pi i}\frac{1-\om}{(1+\om)^3}
\left\{A G(\om)+\om B\tilde{G}(\om)\right\} \nonumber\\
&& \nonumber \\
&=&\frac{A}{2}\left.\left\{\frac{\partial^2}{\partial p^2}
-\frac{\partial}{\partial p}\right\}G(p)\right|_{p=-1}
+\frac{B}{2}\left.\left\{3\frac{\partial}{\partial p}
-\frac{\partial^2}{\partial p^2}-1\right\}\tilde{G}(p)\right|_{p=-1}
\nonumber\\
&&\nonumber \\
&=&\frac{1}{2+n}\frac{a^2}{(1-a^2)^2}\left\{2+e^2-\alpha^2+2\tilde{\alpha}e
\right\}.
\label{susc}
\eea
\subsection{The critical behaviour \label{critical} }
As argued earlier our model becomes singular as $a\rightarrow 0$ (cf.\
to sections~\ref{saddle} and~\ref{O(n)}). Below we will investigate
the nature of the critical behaviour associated with this
singularity. In analogy with what was the case for the ordinary $O(n)$
model on a random lattice the present model only has a well defined
scaling behaviour as $a\rightarrow 0$ if $n\in [-2,2]$ and we will
restrict ourselves to considering this range of $n$ values. One might
also try to look for a critical point associated with $a\rightarrow 1$
(cf.\ to equation~\rf{Bfin}), i.e. with the physical and the
unphysical cut degenerating to the two points $+1$ and $-1$. Due to
the analogy with 
the Penner potential~\cite{Penner} one might expect that having $a=1$
(apart from at $c=0$) is possible only for a particular value of $n$. (If
the analogy were perfect it would be $n=1$). However, we find that the
equation $a=1$ only has the trivial solution $c=0$ regardless of the
value of $n$.
\subsubsection{The case $n \in ]-2,2[$ }
Let us consider the singular behaviour which occurs as $a\rightarrow
0$. First, let us fix $n$ and determine the critical value of
$T$ as a function of $n$. By analysing the $k\rightarrow 0$ limit of
the various elliptic functions which enter the
equation~\rf{Bfin} one concludes that in the limit $a\rightarrow 0$
the dominant term in the curly bracket is $\tilde{\alpha}^2$
and that
\beq
\tilde{\alpha}\sim
i\nu\,\frac{1}{a},
\hspace{0.7cm}\mbox{as}\hspace{0.7cm}a\rightarrow 0.
\eeq
Hence the critical value, $T_*$ of $T$ is 
given by
\beq
(2-n)-\frac{2\nu^2}{T_*}=0
\eeq
or
\beq
T_*=\frac{\nu^2}{2\sin^2(\nu\pi/2)}.
\label{c*}
\eeq
In particular we see that $T_*$ is always positive and greater than
$2/\pi^2$. For $n=1$ we get $T_*=\frac{2}{9}$. In reference~\cite{Cic93} a
numerical determination of this quantity gave $\frac{1}{T_*}=4.504$.

The next to leading order term in the curly bracket in equation~\rf{Bfin}
comes from the term $2\tilde{e}\alpha$ which behaves as
\beq
2\tilde{e}\alpha \sim -\frac{4}{a^2}\,q^{(1-\nu)/2}\sim 
-\frac{4}{a^2}\left(\frac{a^2}{4}\right)^{1-\nu},
\hspace{0.7cm} a\rightarrow 0.
\label{ealpha}
\eeq
{}From this we conclude that
\beq
T_*-T\sim a^{2-2\nu}.
\label{ccstar}
\eeq
Now, let us take a look at the expression~\rf{susc} for $\frac{d}{dT}T^3
\frac{dF}{dT}$.
Here the leading order contribution comes from the term
$\alpha^2$ and is of order
$a^0$. The next to leading order term comes from $2\tilde{\alpha}e$
and is of order $a^{2\nu}$ (cf.\ to equation~\rf{ealpha}).
Bearing in mind the relation~\rf{U(T)} 
we get using~\rf{ccstar}
\beq
U(T)\sim (T_*-T)^{\frac{\nu}{1-\nu}}.
\eeq
This means that
\beq
\gamma_{str}=-\frac{\nu}{1-\nu}.
\eeq

\subsubsection{The cases $n=\pm 2$}

For $n=\pm 2$ the relations~\rf{Bfin} and~\rf{susc} contain
divergent terms. However, the limits $n\rightarrow \pm 2$ of these
relations are well defined. 
\vspace{0.5cm}
\newline
{\bf The case $n=+2$:}
Taking the limit $n\rightarrow 2$ in 
\rf{Bfin} one arrives at the following equation
\beq
\pi^2 -\frac{2}{(1-a^2)^2}\frac{1}{T}\left\{
\left(E'+a^2 K'\right)^2- a^2\left(1+a^2\right)^2 K'^2\right\}=0.
\eeq 
This reproduces the result~\rf{c*} that $T_*=2/\pi^2$ for $n=2$. 
In the limit
$a\rightarrow 0$ the next to leading order contribution in the curly bracket
comes from the term $(aK')^2$ which behaves as $(a\log a)^2$. This gives
\beq
T_*-T \sim \left(a\log a\right)^2
\label{cscaln=2}
\eeq
Furthermore, in the limit $n\rightarrow 2$ the
relation~\rf{susc}
reads
\beq
\frac{d}{dT}\, T^3 \frac{dF}{dT}
=\frac{1}{4\left(1-a^2\right)^2}
\left\{
(1+a^2)^2-4\,\frac{E'}{K'}\right\}.
\eeq
Letting $a\rightarrow 0$ we get
\beq
\frac{d}{dT}\, T^3\frac{dF}{dT}\sim \frac{1}{\log a}.
\label{suscaln=2}
\eeq
The results~\rf{cscaln=2} and~\rf{suscaln=2} coincide with those for
the ordinary $O(2)$ model on a random lattice.
\vspace{0.5cm}
\newline
{\bf The case $n=-2$:}
For $n=-2$ the relation~\rf{Bfin} reduces to
\beq
2-\frac{1}{(1-a^2)^2}\frac{1}{T}\left\{(1+a^2)^2-4\frac{E'}{K'}\right\}=0 
\eeq
which in accordance with~\rf{c*} gives that
$T_*=\frac{1}{2}$. Furthermore it follows that in the limit
$a\rightarrow 0$ 
\beq
T_*-T \sim -\frac{1}{\log a}.
\label{cscaln=-2}
\eeq
The relation~\rf{susc} takes the following form when $n=-2$
\beq
\frac{d}{dT}\, T^3\frac{dF}{dT}=
\frac{1}{(1-a^2)^2}\frac{1}{\pi^2}\left\{
\left(E'+a^2K'\right)^2-a^2(1+a^2)^2K'^2\right\}.
\eeq
In the limit $a\rightarrow 0$ we find
\beq
\frac{d}{dT}T^3\frac{dF}{dT}\sim -\left(a\log a\right)^2.
\label{suscaln=-2}
\eeq
We note that the results~\rf{cscaln=-2} and~\rf{suscaln=-2} do not
coincide with those of the ordinary $O(-2)$ model on a random lattice
which (for gaussian potential) does not have any singular points.

\newsection{Conclusion and outlook \label{conclusion} }
We have solved exactly a hermitian $(n+1)$-matrix model with plaquette
interaction. For $n\in ]-2,2]$ the model was shown to belong to the
same universality class as the $O(n)$ model on a random lattice. 
In particular this result confirms the speculation of
reference~\cite{Cic93} that the critical point of the model~\rf{three}
describes the same physics as the critical point of the $O(1)$ model
on a random lattice.
Using
equation~\rf{susc} it is easy to see that the plaquette 
model has no singular
points (with $T$ finite)
for $n<-2$ and that for $n>2$ the points given by $\bar{\nu} K'=2mK$,
where $\nu=i\bar{\nu}$, are
singular. We expect that in analogy with the ordinary $O(n)$ model,
the solution of the plaquette model breaks down at the first of these
points, $\bar{\nu} K'=2K$, and that the critical index $\gamma_{str}$ takes the
value $+\frac{1}{2}$  at this singularity. Although our model is much
simpler than general lattice gauge models and matrix models generating
Meander numbers our results may be taken as an indication that
elliptic functions might provide a convenient parametrization of such
models.

 Our solution of the plaquette model
contains the solution of a certain three colour problem on a random 
lattice~\cite{Cic93}.
The classical three colour problem due to Baxter~\cite{Baxter}
consists in enumerating all possible ways of colouring
with three different colours the links of a 2D regular three
coordinated lattice in such a way that no two links which meet at the
same vertex carry the same colour. The problem can also be understood
as the problem of counting all possible foldings of the 2D regular
triangulated lattice~\cite{DG94}. 
Obviously the partition function~\rf{three} (with
$V(A)=\frac{1}{2}A^2$) generates {\it random}
lattices with links of three different colours where no two links
radiating from the same vertex have the same colour. Due to the matrix
nature of the fields, however, in the present
 case the cyclic order of the three
colours around a vertex will always be the same. In order to lift this
constraint we would have to introduce two interaction vertices, $\tr
ABC$ and $\tr ACB$. The quartic interaction term in the resulting two
matrix model would then look like $c\left(\tr ABAB +\tr
A^2B^2\right)$~\cite{Cic93}. 
Unfortunately an exact solution of a model with this type of
interaction is still lacking. Let us mention in this connection that a
somewhat similar interaction term,
namely $c\left(\tr ABAB+ 2 \tr A^2B^2\right)$ appears in a matrix model
describing an Ising spin system living on the vertices of a randomly
 quadrangulated surface~\cite{Joh93}.

\vspace{12pt}
\noindent
{\bf Acknowledgements}\hspace{0.3cm}We thank J.\ Ambj\o rn, 
D.\ Johnston, J.\ Jurkiewicz and
 Yu.\ Makeenko  for interesting and useful discussions.
 One of us (L.Ch.) is grateful to the Niels Bohr Institute for hospitality 
 during his visit to NBI where this work was started.

\setcounter{section}{0}
\appendix{The Virasoro constraints}
Like the $O(n)$ model on a random lattice our model can be understood
as a deformation of the one-matrix (n=0) model and obeys a set of
Virasoro constraints obtainable from those of the $n=0$ model by a
canonical transformation~\cite{Kos95}. Let us rewrite~\rf{Aint} as
\beq
Z_n=\int dM \exp \left\{\tr \sum_{i=1}^{\infty}t_i M^i\right\}
\exp\left\{\frac{n}{2}\sum_{k=1}^{\infty}\frac{1}{k}
\left(\tr M^k\right)^2\right\}.
\eeq
Introducing a differential operator, ${\bf H}$, by
\beq
{\bf H}=\frac{n}{2}\sum_{k=1}^{\infty}\frac{1}{k}\left(\frac{\partial}{\partial
t_k}\right)^2 
\eeq
we have
\beq
Z_n=e^{\bf H}Z_0.
\eeq
The $n=0$ model obeys the Virasoro constraints $L_m Z_0=0$, $m\geq -1$
where
\beq
L_m=\sum_{k=0}^m\frac{\partial}{\partial
t_k}\frac{\partial}{\partial t_{m-k}}+\sum_{k=0}^{\infty} kt_k
\frac{\partial}{\partial t_{m+k}}
\eeq
and
\beq
\frac{\partial}{\partial t_0}Z_0=NZ_0.
\eeq
The general model obeys the Virasoro constraints $\tilde{L}_m Z_n=0$,
$m\geq -1$ with
\beq
\tilde{L}_m=e^{\bf H}L_m e^{\bf -H}=
L_m+n\sum_{k=0}^{\infty}\frac{\partial}{\partial
t_k}\frac{\partial}{\partial t_{m+k}}.
\eeq

\end{document}